\documentstyle[12pt]{article}

\def\a{\alpha}
\def\b{\beta}
\def\d{\delta}
\def\e{\epsilon}

\def\f{\phi}

\def\g{\gamma}

\def\k{\kappa}
\def\l{\lambda}

\def\q{\theta}

\def\s{\sigma}

\def\Ld{\Lambda}

\def\Q{\Theta}

\def\cs{{\cal S}}

\def\inbar{\vrule height1.5ex width.4ptdepth0pt}
\def\rlx{\relax\leavevmode}
\def\I{\leavevmode
\hbox{\small1\kern-3.8pt\normalsize1}}
\def\openone{\leavevmode
\hbox{\small1\kern-3.3pt\normalsize1}}
\def\Ione{\rlx{\rm 1\kern-2.7pt l}}
\font\cmss=cmss10
\font\cmsss=cmss10 at 7pt
\def\ZZ{\rlx\leavevmode \ifmmode\mathchoice
{\hbox{\cmssZ\kern-.4em Z}} {\hbox{\cmss Z\kern-.4em Z}}
{\lower.9pt\hbox{\cmsss Z\kern-.36em Z}}
{\lower1.2pt\hbox{\cmsssZ\kern-.36em Z}}
\else{\cmss Z\kern-.4em Z}\fi}
\def\Ik{\rlx{\rm I\kern-.18em k}}
\def\IC{\rlx\leavevmode
\ifmmode\mathchoice
{\hbox{\kern.33em\inbar\kern-.3em{\rm C}}}
{\hbox{\kern.33em\inbar\kern-.3em{\rm C}}}
{\hbox{\kern.28em\sinbar\kern-.25em{\rm C}}}
{\hbox{\kern.25em\ssinbar\kern-.22em{\rm C}}}
\else{\hbox{\kern.3em\inbar\kern-.3em{\rm C}}}\fi}
\def\IP{\rlx{\rmI\kern-.18em P}}
\def\IR{\rlx{\rm I\kern-.18em R}}
\def\IN{\rlx{\rm I\kern-.20em N}}

\newcommand{\ol}\overline
\newcommand{\ul}\underline
\newcommand{\ti}\tilde
\newcommand{\wt}\widetilde
\newcommand{\wh}\widehat
\newcommand{\bv}\breve
\newcommand{\dg}\dagger

\newcommand{\be}{\begin{equation}}
\newcommand{\ee}{\end{equation}}
\newcommand{\bl}{\begin{eqnarray}&}
\newcommand{\el}{&\end{eqnarray}}
\newcommand{\bq}{\begin{eqnarray}}
\newcommand{\eq}{\end{eqnarray}}

\newcommand{\ov}{\overline}
\newcommand{\pa}{\partial}

\begin{document}
\title{\bf A New Version of Dirac's \AE ther and Its 
Cosmological Applications}
\author{
Marcelo Carvalho \thanks{ {\sc e-mail:
marcelo\underline
{\hspace{2mm}}carv@hotmail.com} } \\
\normalsize{Waseda University} \\
\normalsize{Department of Mathematics }\\
\normalsize{3-4-1 Okubo, Shinjuku-ku, Tokyo 169, Japan}
\\
and 
\\
A. L. Oliveira \thanks{ {\sc e-mail: alexandr@ov.ufrj.br}}\\
\normalsize{Universidade Federal do Rio de Janeiro} \\
\normalsize{Grupo de F\'{\i}sica e Astrof\'{\i}sica Relativista (GFAR)}  \\
\normalsize{Observat\'{o}rio do Valongo} \\
\normalsize{20080-090 Rio de Janeiro -- RJ, Brazil}  \vspace{3mm}
\\
   }
\date{}
\maketitle
\eject
\begin{abstract}
We propose a new formulation for the \AE ther of Dirac based 
on a lagrangian approach. We analyse how the presence of a 
particular self-interaction term in the lagrangian lead us to a description
of the \ae ther as being a medium with conductivity which is governed 
by macroscopic Maxwell equations with a polarization tensor 
$M_{\a\b}$ depending on the vector potential. These results are 
then applied to the analysis of the amplification of the primordial magnetic
induction in a curved background of Friedmann's geometry.
\noindent
 \end{abstract} 

\section{Introduction}
The original version of the Dirac's \AE ther was presented long 
ago as a result following his formulation of 
the so-called ``New Electrodynamics"  (NE) \cite{Dirac2}. In the 
NE, Dirac has adopted a radical approach for the discussion of 
a classical electron in the presence of an electromagnetic field. The 
main idea was to use the spurious degrees of freedom associated 
to the gauge potential to describe the electron. In \cite{Dirac2} Dirac 
showed how this could be done by a judicious choice of gauge 
condition, e.g. $A^{2}= k^{2}$. In fact, introducing this as a 
gauge fixing term in the Lagrangian 
{\footnote{With $\l$ a lagrange multiplier and k a constant.}}
$\mbox{{\it L}} = -\frac{1}{4} 
F^{2}+ \frac{\lambda}{2}(A^{2} - 
\kappa^2), $ we find $\pa_{\nu}F^{\nu\mu}= J^{\mu}\equiv \l A^{\mu}$. 
Therefore, we obtain a four current depending on $A_{\mu}$ that is suitable 
for the description of a classical electron interacting with an 
electromagnetic field. In this approach we observe that the gauge 
condition doesn't intend to eliminate spurious degrees of freedom 
of the gauge potential. Instead of that, it acquires 
a deep physical meaning as the condition that allow us to describe 
the right physics without having to introduce any extra fields.

One of the striking consequences of these results were discussed 
in a subsequent paper of Dirac \cite{Dirac} and led to the \AE ther 
model. In fact, by using quantum mechanical arguments Dirac 
argued it would be possible to consider an \ae ther provided we
interpret its four velocity $v$ at each point as a quantity subjected to 
uncertainty relations. Then, the \ae ther four velocity 
instead of being a well-defined quantity, would rather be probabilisticaly 
distributed over a range of values. The wave function describing the 
\ae ther would be such as to assure a velocity distribution with all 
velocities being equally probable. Admitting the \ae ther velocity as 
defining a point in a hyperboloid with equation $ v^{2}_{0} - \vec{v}^{2}=1,\; 
v_{0} > 0$, Dirac could relate it to the gauge potential (satisfying $A^2 = 
k^{2}$) by $\frac{1}{k} A_{\mu}=v_{\mu}$ and from the interpretation given 
to the NE, he concluded that $v$ would be the velocity an electric charge 
would flow if placed in the \ae ther. The four velocity $v$, being defined 
through all points of spacetime, justify its interpretation as a meaningful 
physical quantity, the \ae ther velocity. 

Following Dirac, many authors have tried to develop a realistic model 
for the Dirac's \AE ther by considering the \ae ther as a vacuum with 
conductivity $\sigma \approx 10^{-13}$/s \cite{Vigier,Kar}. They have also 
found the photon as having a non-null mass 
$m_{\gamma}=\sigma h/c^2\approx 10^{-48}$g (which is 
below laboratories estimates for the limit of the photon mass). 
In these models, the basic equations for the \ae ther were the Maxwell equations 
in a conducting medium without density of charge and current. The intensity of the
electric field $\vec{E}$, and the magnetic induction $\vec{B}$ 
\footnote{Here we will also refer to the intensity of the electric field $\vec{E}$
and the magnetic induction $\vec{B}$ as simply electric and magnetic field.}
are related to the electric displacement $\vec{D}$,
and the magnetic intensity $\vec{H}$ by the relations $\vec{D}=\e \vec{E},\; 
\vec{H}=\frac{1}{\mu}\vec{B}$ which suppose the \ae ther as a homogeneous 
medium polarized isotropically and linearly. Besides that, it is implicit in 
\cite{Vigier,Kar} that all their analysis is considered in a flat spacetime and 
from a reference frame at rest relative to the \ae ther.

Our present work is a continuation of an investigation initiated in a
 preceding paper \cite{Alex3} where we have studied the amplification 
of the cosmic magnetic induction for the Proca electromagnetic field 
in the Dirac's \AE ther. There the equations used were the Proca's 
equations in a conducting media. Here, we wish to study the Dirac's 
model from another point of view, extending his original idea but 
maintaining the physical motivation for the application in a curved 
background. One of our interests is to analyse how the primordial 
cosmic magnetic induction has been amplified, as the Universe evolves, 
for a model that is neither Proca nor the purely Maxwell electrodynamics 
(but that includes this last one as a particular case). In the literature of the 
Dirac's \AE ther, none of the models presented in \cite{Vigier, Kar, 
Sudarshan} discusses a possible framework for the study of 
electromagnetic phenomena in the large scale structure. They all 
take for granted the basic Maxwell theory. Therefore, any application 
to a large scale scenario is expected to be modeled within the 
Maxwell formulation. Since it is unknown if under the influence of a 
curved background the electromagnetic field presents a different 
behaviour than that foreseen by the Maxwell theory, it is reasonable 
to test alternative approaches that may provide a suitable description 
for the electromagnetic field at the large scale. Our model intends to 
offer such an alternative.  

Although we are not concerned with the relevant question
about the structure of the \ae ther as a conducting medium
\cite{Sudarshan}, it is implict in our formulation that
there is an inertial reference frame in which the \ae ther is at rest and
consequently the field of velocities of the points of the \ae ther write
as $v^{\mu}_{\ae ther}=(1,0,0,0)$. We refer it as the \ae ther frame. 
The main point of our research is based on the proposition of an action 
for the \AE ther of Dirac which may be able to reproduce some characteristics of 
previous models. We propose then the action 
$\cs = \int dx \Bigl(-\frac{1}{4} F^{2} + \s v_{\a}F^{\a\mu} A_{\mu} \Bigr)$ 
with $v$ being the  \ae ther's velocity relative to a generic observer (inertial or not).
For inertial observers, $v^{\mu}=\Ld^{\mu}_{\nu}v^{\nu}_{\ae ther}=\Ld^{\mu}_{0}$
\footnote{with $\Ld$ the Lorentz transformation relating the observer and the \ae ther frame.}
is still constant and we obtain from the equations of motion a 
four-current in the form $J^{\mu}=-\s v^{\mu} \pa . A + 
\s v_{\nu}\pa^{\mu} A^{\nu}$, that is understood as being 
induced in the \ae ther by the presence of the electromagnetic 
field. Therefore, it defines a polarization tensor $M_{\a\b}$ 
from which we obtain the vectors of polarization and magnetization 
of the medium. We will see in section {\bf 2.4} that in the \ae ther
reference frame the form assumed by this 4-current allow us to define the 
electric displacement vector as $\vec{D}=\vec{E} +\s \vec{A}$ while 
the magnetic intensity coincides with the magnetic field, 
$\vec{H}=\vec{B}$, with the resulting equations being similar in 
form with the macroscopic Maxwell equations in a medium. This 
coincides with the results of \cite{Vigier, Kar} but with a particular 
difference: $\vec{D} \neq \e \vec{E}$. This shows that in our model 
the \ae ther cannot be thought of as an isotropic medium. Moreover, in a generic 
reference frame moving relative to the \ae ther, the vectors of electric 
displacement $\vec{D}$ and magnetic intensity $\vec{H}$ will depend 
on $v$. Our treatment offers an extension of the results of \cite{Vigier, Kar} 
in the sense that our equations incorporate the situation of reference 
frames that are not at rest relative to the \ae ther. 

In our approach the basic electromagnetic
fields are the vectors $\vec{E}$ and $\vec{B}$ which are
components of the tensor $F_{\mu\nu}\equiv
\pa_{\mu}A_{\nu}-\pa_{\nu}A_{\mu}$. The equations of motion
(\ref{m1}) are determined directly from our action, with the Jacobi
identity $F_{\mu\nu,\a}+ F_{\nu\a,\mu}+ F_{\a\mu,\nu}=0$
completing the set of equations involving the electromagnetic field.
In the traditional formulation of classical electrodynamics
the non-homogeneous Maxwell equations in a conducting
medium are obtained introducing a second tensor
$H^{\mu\nu}$ with components $\vec{D}$ and $\vec{H}$ which satisfy
$\pa_{\mu}H^{\mu\nu}= -j^{\nu}$. Compared to our formalism we
don't have to introduce by hand any tensor $H^{\mu\nu}$.
Here it is the interaction term $v_{\a}F^{\a\mu}A_{\mu}$ that
forces the appearance of additional terms (depending on $A_{\mu}$) in the
equations for $\vec{E}$ and $\vec{B}$ which allow us to identify as $\vec{D}$
and $\vec{H}$.
Both the classical electrodynamics in vacuo (\ref{free1}-\ref{free4})
and in conducting media (\ref{ether1}-\ref{ether4}) are obtained as
a particular case of our model for an observer that is at rest
relative to the \ae ther. For observers that move slowly relative
to the \ae ther and/or for phenomena in which the \ae ther
conductivity is sufficiently small we are expected to have the interaction
term giving a small contribution to the Maxwell equations and
therefore the model may qualitatively agree with the same
predictions of the traditional formulation of classical
electrodynamics. 

\begin{sloppypar}
This work is organized as follows. 
Considering an observer moving with constant velocity relative to the \ae ther,
in Section {\bf  2.1} we discuss the lagrangian of the model, identify the 4-current and a 
constraint that appears associated to its conservation; in Section {\bf 2.2} we analyse a
global gauge invariance of the action, we show that in the \ae ther
frame the conservation of the Noether charge coincides with the
Gauss law, we interpret the interaction term
$v_{\a}F^{\a\mu}A_{\mu}$ as $\ti J \cdot A$ with
${\ti J}^{\mu}\equiv v_{\a}F^{\a\mu}$ a conserved 4-current,
and we propose a mass term that preserves the global symmetry; in
Section {\bf 2.3} we show that our action admits a local gauge
invariance; in Section {\bf 2.4} we show that in the \ae ther frame we 
obtain macroscopic Maxwell equations with $\vec{D}=\vec{E} + \s \vec{A}$ and 
$\vec{H}=\vec{B}$. Considering the case of a curved background we will
have, in general, $v\neq cte$; in Section {\bf 3} we
establish the material relations involving the fields; in Section {\bf 4} 
the equations of the model in a cosmological background are solved 
and the solutions are discussed.  The last section 
is devoted to concluding remarks.
\end{sloppypar}

\subsection*{Notations For Flat Space-time}
\bq
\eta_{\mu \nu}&=&{\mbox{diag}}(+,-,-,-) \nonumber \\
x^{\mu}&\equiv& (x^{0},x^{i})\doteq (t,\vec{x}),\;\;\;
x_{\mu}\equiv (x_{0},x_{i})\doteq (t,-\vec{x}) 
\nonumber\\
\pa^{\mu}&\equiv& (\pa^{0},\pa^{i})\doteq (\frac{\pa}{\pa t}, -\vec{\nabla}),\;\;\;
\pa_{\mu}\equiv (\pa_{0},\pa_{i})\doteq (\frac{\pa}{\pa t}, \vec{\nabla}) \nonumber \\
A^{\mu}&\equiv& (A^{0},A^{i})\doteq (\f,\vec{A}),\;\;\; 
A_{\mu}\equiv (A_{0},A_{i})\doteq (\f,-\vec{A}) \nonumber \\
F_{0i}&=& \vec{E}_{i},\;\;F_{ij}=-\e_{ijk}\vec{B}_{k},\;\;\;
F^{0i}=-F_{0i}=-\vec{E}_{i},\;\;\;
F^{ij}=F_{ij}=-\e_{ijk}\vec{B}_{k} \nonumber
\eq

\section{The Lagrangian of the Dirac's \AE ther and Its Invariances}
\subsection{Equations of Motion}
Let us consider a certain reference system moving with 
a constant 4-velocity relative to the \ae ther reference frame. 
We take for action
\be
\cs = \int dx \Bigl(-\frac{1}{4} F^{2} +{\ti J}\cdot A \Bigr)
\label{l}
\ee
with 
\be
{\ti J}^{\mu} \doteq \s v_{\a}F^{\a\mu} \;.
\label{j1}
\ee
In these expressions we have
$F_{\mu\nu}=\pa_{\mu}A_{\nu}-\pa_{\nu}A_{\mu}$, and the field
$A$ is understood as a 1-form defined in a flat spacetime manifold. 
The (constant) parameter $\s$ is associated to the \ae ther 
conductivity and $\ti J$ will be shown to be a conserved quantity. 
Here, the term ${\ti J}\cdot A$ defines an interaction of the gauge field with itself. 

The equation of motion for $A_{\mu}$ has the form
\be
\pa_{\nu}F^{\nu\mu}+\s v^{\mu} \pa . A - \s v_{\nu}\pa^{\mu} A^{\nu}=0
\label{m1}
\ee
and it assumes the same form as the Maxwell equations in the presence 
of a source 
\be
\pa_{\nu}F^{\nu\mu}\equiv J^{\mu}
\label{m2}
\ee
provided we identify 
\be
J^{\mu}=-\s v^{\mu} \pa . A + \s v_{\nu}\pa^{\mu} A^{\nu}
\label{j2}
\ee
with a conserved 4-current. This interpretation follows the same idea of
Dirac's NE \cite{Dirac2} in which the term
$j^{\mu}\doteq \l A^{\mu}$ was interpreted as a 4-current. However, in Dirac's 
formalism the appearance of this 4-current originates from the introduction of a
gauge fixing term 
$\frac{1}{2}\l (A^{2} - k^{2})$ in the action while in our approach $J^{\mu}$, 
given in (\ref{j2}), arises from the presence of the interaction term ${\ti J}.A$.

Now, taking the divergence of (\ref{m2}) we obtain
$0=\pa_{\mu} J^{\mu}=\s v_{\mu}(\Box A^{\mu}- \pa^{\mu} \pa . A)
=\s v_{\mu}\pa_{\nu}F^{\nu\mu} =\s v_{\mu}J^{\mu}=\s^{2} (-v^{2}\pa . A +
v_{\a}v_{\b}\pa^{\a}A^{\b}) $, i.e
\be
\pa . A = \frac{v_{\a}v_{\b}}{v^{2}} \pa^{\a}A^{\b}\;.
\label{condition1}
\ee
This constraint is a new feature of our model and since it involves
the divergence of $A$ it resembles a kind of gauge condition. 
Nonetheless, its origin is independent of any local symmetry
of the action. In section 2.3 we will analyse how this condition
combines with a local symmetry of the action imposing 
certain conditions to be satisfied by the gauge parameter.

\subsection{Global Gauge Invariance}
The first invariance of the action we want to analize is the one 
defined by a transformation parametrized by a global 
parameter $\l\; (\pa_{\mu} \l=0)$ and having the form
\be
A_{\mu} \rightarrow A'_{\mu}= A_{\mu} + \l v_{\mu} \;.
\label{global}
\ee
Associated to this symmetry we have the following Noether 
current
\be
\Q^{\mu}=F^{\mu\nu}v_{\nu} -\s v^{\mu}A.v +\s v^{2} A^{\mu}\equiv 
\Q^{\mu\nu}v_{\nu}
\label{noether}
\ee
with 
\be
\Q^{\mu\nu}\doteq F^{\mu\nu} -\s v^{\mu}A^{\nu} + \s v^{\nu}A^{\mu}\;.
\label{Q}
\ee
In section {\bf{3}} the quantitity $\Q^{\mu\nu}$ will be 
interpreted as the tensor $H^{\mu\nu}$, which in the 
\ae ther frame becomes $H^{\mu\nu}=(\vec{D},\vec{H})$. In this general 
case,  it will be clear how $H^{\mu\nu}$ depends on the 
properties of the medium and on the 4-velocity $v$. 
In a system that is at rest relative to the \ae ther we have
\be
\Q^{\mu}=(0, \vec{E} + \s\vec{A})
\ee
and condition (\ref{condition1}) gives 
$\vec{\nabla} . \vec{A}=0$. Therefore,  the conservation 
equation of $\Q^{\mu}$ let us with 
\be
\vec{\nabla} . \vec{E} =0 \;.
\ee 
\noindent
In our model, there is another conserved current that has 
the form
\be
{\hat J}^{\mu}= -\s v^{\mu} \pa . A + \s v . \pa A^{\mu}
\ee
and from which we obtain ${\ti J}={\hat J}- J$. Then, 
conservation of ${\ti J}$ follows immediately as the 
difference of two conserved currents. Equivalently, 
from (\ref{Q}) we can also think of ${\ti J}^{\nu}$ as 
originating from the divergence of $\Q^{\mu\nu}$, 
\bq
\pa_{\mu}\Q^{\mu\nu}=-{\ti J}^{\nu}\;. 
\label{corrente}
\eq
In the classical formulation of Electrodynamics in conducting media
the non-homogeneous Maxwell equations are written covariantly as
\bq
\pa_{\mu}H^{\mu\nu}= - j^{\nu}_{ext}
\label{h}
\eq
with the tensor $H^{\mu\nu}$ having $\vec{D}$ and $\vec{H}$
as its components. In our model $\Q^{\mu\nu}$ (\ref{Q}) generalizes
the tensor $H^{\mu\nu}$ and equation (\ref{corrente}) corresponds
to (\ref{h}). Then this allow us to interpret ${\ti J}^{\mu}$ as the 
corresponding 4-current of our model, in much the same way as Dirac
interpreted $j_{\mu}=\l A_{\mu}$ as a 4-current in \cite{Dirac2}.
The interaction term ${\ti J}\cdot A$ in our action (\ref{l}) parallels
then the same interaction term of the usual electrodynamics.

The global symmetry (\ref{global}) is a new feature of 
our model that has no similar counterpart in the usual 
Maxwell formulation. It is also possible to add a mass 
term to the action (\ref{l}) that perserves this global 
symmetry. In fact, the action 
\be
\cs = \int dx \Bigl(-\frac{1}{4} F^{2} +{\ti J}\cdot A  - \frac{1}{2}
\s^{2} A_{\mu}(v^{2} g^{\mu\nu} - v^{\mu}v^{\nu})A_{\nu}\Bigr)
\label{l2}
\ee
is invariant by (\ref{global}). From (\ref{l2}) we obtain the 
following equation
\be
\pa_{\nu}F^{\nu\mu}\equiv \ov{J}^{\mu}\doteq -
\s v^{\mu} \pa . A + \s v_{\nu}\pa^{\mu}A^{\nu}+
\s^{2} (v^{2} g^{\mu\nu}-v^{\mu}v^{\nu})A_{\nu}
\label{mass}
\ee
or equivalently
\be
\Bigl(g^{\mu\nu} ( \Box -\s^{2} v^{2}) - \pa^{\mu}\pa^{\nu} +
\s (v^{\mu}\pa^{\nu} -v^{\nu}\pa^{\mu}) + 
\s^{2} v^{\mu}v^{\nu} \Bigr)A_{\nu}=0\;.
\ee
Here the conservation of the current $\ov{J}^{\mu}$ that 
follows from (\ref{mass}) doesn't produce any constraint 
over $A$. We also notice that associated to the global 
symmetry we have the same conserved 
current (\ref{noether}).

\subsection{Local Gauge Invariance}
The local gauge invariance depends on a parameter $\q(x)$ 
and assumes the usual form
\be
A_{\mu} \rightarrow A'_{\mu}=A_{\mu} + \pa_{\mu}\q \;.
\label{gauge2}
\ee
Here, condition (\ref{condition1}) adds new features to our 
analysis. In fact, let $A'$ and $A$ be two fields related by 
(\ref{gauge2}). Since both field configurations should obey 
(\ref{condition1}) we must have
\be
\pa . A' = \frac{1}{v^{2}}(v.\pa)(v. A') \rightleftharpoons 
\Box \q = \frac{v_{\a}v_{\b}}{v^{2}}
\pa^{\a}\pa^{\b}\q\;.
\label{con}
\ee
Consider that $\pa . A \neq 0$ and let us choose $\q$ such 
that it ensures $\pa . A'=0$. We should then have $\q$ 
satisfying 
\be
\Box \q = -\pa . A \;.
\label{con1}
\ee
This last condition together with the constraints 
(\ref{condition1},\ref{con}) gives 
\be
\pa^{\a}\pa^{\b} \q = -\frac{1}{2}(\pa^{\a}A^{\b} + \pa^{\b}A^{\a})
\label{constraint2}
\ee
that represents a restriction stronger than that shown in (\ref{con1}). 
Equivalently, we can obtain equation (\ref{constraint2}) directly from
$0=\pa . A' = \frac{v_{a}v_{\b}}{v^{2}}\pa^{\a}(A^{\b} 
+ \pa^{\b}\q)$ by using (\ref{condition1}).

\subsection{Electrodynamics in the \AE ther's Reference 
Frame} \label{EER}

Let us analyse our model in the \ae ther reference frame. We have $v=(1,0,0,0)$.
We also suppose the \ae ther as a medium without any given density of charge or 
current.  From (\ref{m2}) we obtain the following equations
\bq
\vec{\nabla}. \vec{E} &=& - \s \vec{\nabla} . \vec{A} 
\label{gauss1}\\
\vec{\nabla} \times  \vec{B} &=& \frac{\pa \vec{E}}{\pa t}
+ \s \frac{\pa \vec{A}}{\pa t} +\s \vec{E}  
\label{ampere}
\eq
to which we add the homogeneous equations
\bq
\vec{\nabla} . \vec{B} &=& 0 
\label{gauss2}\\
\vec{\nabla} \times \vec{E} &=& - \frac{\pa \vec{B}}{\pa t}\;.
\label{faraday}
\eq
Since $\vec{\nabla}. \vec{E} \neq 0$, we have that (\ref{gauss1}) introduces a new feature for the physical vacuum. A similar situation has been observed in the extended electromagnetism of \cite{Lehnert}. Essentially, a divergenceless equation for $\vec{E}$ signalizes that the vacuum is not merely an empty space but it is also subjected to become electrically polarized. The presence of
additional terms depending on the potential vector in (\ref{gauss1},
\ref{ampere}) is understood as signalizing the response of the
medium to the presence of the fields $(\vec{E},\vec{B})$, a situation that
resembles the phenomena of polarization and magnetization
of a medium. Therefore, we rewrite (\ref{gauss1}) as
\be
\vec{\nabla} . \vec{D}=0 
\ee
with $\vec{D}\doteq \vec{E} +\s \vec{A} + \vec{\nabla}\times \vec{K}$.
At this point, $\vec{K}$ is an arbitrary vector that can be thought of as playing 
a similar role of a gauge parameter. Now, we rewrite
(\ref{ampere}) as
\be
\vec{\nabla} \times \vec{B} = \frac{\pa \vec{D}}{\pa t} +\s \vec{E} 
- \frac{\pa}{\pa t}\vec{\nabla}\times \vec{K}
\ee
and we choose $\vec{K}$ such that it satisfies
\be
 \frac{\pa}{\pa t}\vec{\nabla}\times \vec{K}=\s \vec{E}\;.
\label{K1}
\ee
Then, using (\ref{faraday}) we obtain 
\be
\frac{\pa}{\pa t} \vec{\nabla} \times 
\Bigl(\s \vec{A} + \vec{\nabla} \times \vec{K}\Bigr)= 0\;.
\label{K2}
\ee
The vector $\vec{K}$ can be further restricted so that 
\be
\s \vec{A}+ \vec{\nabla}\times \vec{K} = 0\;.
\label{K3}
\ee
This gives  
$\vec{E}=\vec{D}$ and $\vec{B}=\vec{H}$ and lead us to a set
of equations
\bq
\vec{\nabla} . \vec{E}&=&0
\label{free1} \\
\vec{\nabla} \times \vec{B} &=& \frac{\pa \vec{E}}{\pa t} \\
\vec{\nabla} . \vec{B} &=& 0 \\
\vec{\nabla} \times \vec{E} &=& - \frac{\pa \vec{B}}{\pa t}
\label{free4}
\eq
that correspond to the Maxwell equations in free space. 
Conditions (\ref{K1},\ref{K3}) can be interpreted as originating from 
the imposition of the temporal gauge. In fact, together they 
imply $\vec{\nabla} A_{0}=0$ that is naturally
satisfied if we put $A_{0}=0$.

It is possible to give another description for our electrodynamics 
without using this vector $\vec{K}$. From (\ref{ampere}) we can 
simply identify $\vec{D}= \vec{E} + \s \vec{A},\;
\vec{H}=\vec{B}$
that lead us to the following equations
\bq
\vec{\nabla} . \vec{D}&=&0 
\label{ether1}\\
\vec{\nabla} \times \vec{B} &=& \frac{\pa \vec{D}}{\pa t} +\s \vec{E} \\
\vec{\nabla} . \vec{B} &=& 0 \\
\vec{\nabla} \times \vec{E} &=& - \frac{\pa \vec{B}}{\pa t}
\label{ether4}
\eq
that coincides with the \ae ther's equations obtained in \cite{Vigier, Kar}.
In the identification $\vec{D}= \vec{E} + \s \vec{A},\; \vec{H}=\vec{B}$ 
the \ae ther behaves like a medium that responds to the presence of the electric
field by creating a polarization $\vec{P}=\s \vec{A}$. We also have
a current $\vec{J}= \s \vec{E}$ which is in agreement with our
supposition of the \ae ther as being a medium with conductivity $\s$.

According to Schwinger's idea of 
structureless vacuum \cite{Schwinger}, an electromagnetic field disturbs 
the vacuum affecting its properties of homogeneity and isotropy. 
This is exactly the situation we have obtained in our model, where
the presence of an electromagnetic field in a vacuum with conductivity $\s$
produces a response of the medium $(\vec{D} \neq \e \vec{E})$
that signalizes its non-isotropy.

\section{Material Relations Involving $H^{\mu\nu}$ and $F_{\mu\nu}$}
In section \ref{EER}, we have obtained the following material relations 
\bq
\vec{D}&=& \vec{E} + \s  \vec{A} 
\label{mat1}\\
\vec{H}&=&\vec{B}
\label{mat2}
\eq
for a flat spacetime and for the case of a reference frame 
at rest relative to the \ae ther. In the case of a curved spacetime
and for a non-inertial reference frame
moving relative to the \ae ther, we are supposed 
to have a more complicated relation between $H$ and $F$. 
Indeed, the general form for the material relations between the tensors 
$H^{\mu\nu}$ and $F_{\mu\nu}$ in a medium that is at rest in {\it{any}} 
reference frame with a metric $g_{\a\b}$ has been established in \cite{volkov} as:
\be
\sqrt{-g}H^{\a\b}=\sqrt{-g} g^{\a\g}g^{\b\k} S^{\mu}_{\g}S^{\nu}_{\k}F_{\mu\nu }
\label{mat3}
\ee
where the tensor $S^{\a}_{\b}$ characterizes the electromagnetic properties of the
medium. For our later purpose, it is convenient to rewrite (\ref{mat3}) as
\bq
\sqrt{-g}H^{\a\b}=\sqrt{-g} g^{\a\g}g^{\b\k} S^{\mu}_{\g\k}A_{\mu}
\label{mat33}
\eq
where the third rank mixed tensor $S^{\mu}_{\g\k}$relates to $S^{\a}_{\b}$ by 
\bq
S^{\mu}_{\g\k}=(S^{\nu}_{\g}S^{\mu}_{\k}-S^{\mu}_{\g}S^{\nu}_{\k})\pa_{\nu}\;.
\label{mat333}
\eq
As an application of (\ref{mat3}) it was shown in
\cite{volkov} that for the vacuum (considered from an inertial reference
frame) the tensor  $S^{\a}_{\b}$ assumes the form $S^{\a}_{\b}=\d^{\a}_{\b}$ and the material 
equations become $\sqrt{-g}H^{\a\b}=\sqrt{-g} g^{\a\mu}g^{\b\nu} F_{\mu\nu}$
which reproduces the usual equations of free electrodynamics in a curved 
background \cite{plebanski,landau}. Also, in the case of a linear isotropic medium that is at rest\
in an inertial reference frame the tensor $S^{\a}_{\b}$ is
given by $S^{0}_{0}=\e \sqrt{\mu},\;S^{1}_{1}=S^{2}_{2}=S^{3}_{3}=\frac{1}{\sqrt{\mu}}$
and we obtain the usual relations $\vec{D}=\e\vec{E}$ and $\vec{H}=\frac{1}{\mu}\vec{B}$.
for the electrodynamics in a medium.

In our model, in order to define the tensor $H^{\a\b}$ for a generic reference frame in a 
curved background and to find a suitable material relation of the
type (\ref{mat33}) we should first follow the procedure of section \ref{EER}
that allow us to define $H^{\mu\nu}$ directly from the equation of motion for $A_{\mu}$.
Explicitly, let us take the equation of motion in a curved background,
\be
 \pa_{\nu}(\sqrt{-g}F^{\nu\mu}-\s\sqrt{-g}v^{\nu}A^{\mu} + \s\sqrt{-g}v^{\mu}A^{\nu})=
J^{\mu}\equiv -\s\sqrt{-g}v_{\nu}F^{\nu\mu}
\ee
then we define 
\bq
\sqrt{-g}H^{\a\b}\doteq \sqrt{-g}(F^{\a\b}-\s v^{\a}A^{\b} + \s v^{\b}A^{\a})\;.
\label{mat4}
\eq
This procedure is equivalent to the introduction of  the antisymmetric polarization tensor
of the medium $M^{\a\b}$ \cite{volkov}
\be
\sqrt{-g}H^{\a\b} = \sqrt{-g} (F^{\a\b} + M^{\a\b}) \rightleftharpoons 
\sqrt{-g}H^{\a\b} = \sqrt{-g} g^{\a\nu}g^{\b\mu}(F_{\nu\mu} + M_{\nu\mu}) 
\label{mat5}
\ee
provided we identify from (\ref{mat4},\ref{mat5})  
\be
M^{\a\b}=-\s v^{\a}A^{\b} + \s v^{\b}A^{\a} \rightleftharpoons 
M_{\a\b}=-\s v_{\a}A_{\b} + \s v_{\b}A_{\a} 
\ee
with $F^{\mu\nu}=g^{\mu\a}g^{\nu\b}F_{\a\b},\;F_{\mu\nu}\equiv
\pa_{\mu}A_{\nu}-\pa_{\nu}A_{\mu},\;M^{\mu\nu}=g^{\mu\a}g^{\nu\b}M_{\a\b}$.
Finally, in order to obtain the material equations for our model we extend
definition (\ref{mat33}) by allowing $S^{\mu}_{\a\b}$ to be a generic operator 
not restricted by (\ref{mat333}) but given by
\be
S^{\mu}_{\a\b}= \d^{\mu}_{\b}(\pa_{\a}-\s v_{\a}) -
\d^{\mu}_{\a}(\pa_{\b}-\s v_{\b})\;.
\label{mat6}
\ee
Here, the tensor $S^{\mu}_{\a\b}$ may contain not only electromagnetic
properties of the medium (as it is the case of (\ref{mat333})) but also
information about the reference frame (implicit in the four-velocity $v_{\a}$).
Adopting the convention $\vec{D}_{i}=\sqrt{-g} H^{i0},\;\vec{H}_{i}=
-\frac{1}{2} \e_{ijk} \sqrt{-g}H^{jk}$ \cite{volkov} and considering a flat
spacetime we obtain (in the \ae ther frame) from (\ref{mat33},\ref{mat6}) the 
relations given in (\ref{mat1},\ref{mat2}).

{\raggedright
\section{The Dirac's \AE ther in a Curved Background} }  \label{CB}
\subsection{Some Questions on the Cosmic Magnetic Induction} \label{SQC}
The origin of the cosmic magnetic induction (CMI) is still unknown.
In fact, until now no theory has completely succeeded on explaining 
the evolution of the CMI, from its generation in the early universe to
our present time in galaxies, stars and probably in the extracluster 
medium. The standard dynamo action, for instance,  has left many questions 
without answer \cite{Rubinstein, Parker}. Recently, one special type of 
mechanism for the amplification of the CMI has been studied in 
\cite{Alex3} where the amplification was understood as being 
caused by the expansion of the cosmological background. This 
new kind of theoretical preview was called {\it geometric amplification} 
(GA) because the only agent responsible for this effect was the scale 
geometric factor $R(t)$. One of the advantages of this approach is that 
the amplification can be an alternative for the standard mechanism recently 
contested in \cite{Rubinstein}.

The radio emission of very distant cosmic objects determines
an upper limit for the intensity of the CMI. According to the literature,
in the extra-galatic medium the CMI is supposed to have an intensity 
less than $10^{-13}$T \footnote{Other models have considered a generation of 
a primordial CMI of magnitude $\approx 10^{-21}$T \cite{Rubinstein}. }.

\subsection{The Model in the Friedmann's Geometries } \label{MFCB}
Considering these ideas, we will now apply our model to a curved 
background. We will be particularly interested on finding solutions for 
the CMI that may provide an explanation for the amplification of this 
field from the initial value of $\approx 10^{-19}$T to the suggested actual value of 
$\approx 10^{-13}$T. Here, we will assume that the conductivity of 
Dirac's \AE ther is $\approx 10^{-19}$/s. 

Let us take as coordinate system the cylindrical coordinates of Schr\"odinger $x^ {\mu } = 
(t;\,  \rho ,\,  \phi ,\,  \zeta )$ in a Friedmann cosmological background 
minimally coupled with the electromagnetic field. The metric tensor 
in all three Friedmann geometries is written  as
\begin{eqnarray} \label{ds2-d}
g_{\mu \nu} =  diag\left[R^{2}(t)(1\,; -1 \,, -u^2 \, , - w^2 )\right],
\end{eqnarray}
where the functions $u(\rho)$ and $w(\rho)$ define the type of three-geometry 
(with constant curvature $k_{c}$) under consideration. These functions and the
different scale factors are shown in table 1. 
\begin{table}[h]
\caption{The functions $u$, $w$ and R(t) where $\alpha = 10^{26}$m. }
\label{tabela1}
\begin{center}
\begin{tabular}{c c c} \hline
              $u$ and $w$ & $k_{c}$&$ R(t)$
  \\ \hline \hline
  $u = \rho , w = 1 $  &  $ 0 $ &$(\alpha/2) t^{2}$
  \\   \hline
  $u = \sin\rho, w = \cos\rho$ &  $+1$ &$\alpha (1 - \cos t)$
 \\   \hline 
$u = \sinh\rho, w = \cosh\rho$ & $-1$  & $\alpha (\cosh t -1) $
 \\ \hline
\end{tabular}
\end{center}
\end{table}
We assume here the same time dependent 4-potential with cylindrical symmetry \cite{Alex3} given by
\begin{eqnarray} \label{Amu}
A^{\mu}  =  (0;\,  0,\,  1,\,  0)f(t)/R^2(t),
\end{eqnarray}
where $f(t)$ is a function to be determined by the field equations. 
The non-zero components of the field strength $F_{\mu \nu}$ 
are $F_{02}$ and $F_{12}$. Therefore, for the orthonormal basis 
we have chosen, the non-null components of the fields ${\bf E}$ 
and $c{\bf B}$ are $E_\phi = - \dot{f} u^2/R^2$, $cB_\zeta = 2 f u\;u'/R^2$
\footnote{The dot means $d/dt$, the prime means $d/d\rho$, and $c$, in this 
section, is written explicitly as the light velocity.}.
To study how the expansion of the Universe influences the
time evolution of the electromagnetic field, it is appropriate to define 
the time dependent quantities ${\cal E}(t)$ $\equiv$ $|{\bf E}/u| = | \dot{f}|/R^2$ 
and ${\cal B}(t)$ $\equiv$ $|{\bf B}/u'| = 2 |f |/(cR^2)$ that for simplicity we will
also refer as the electric field and magnetic induction. 

In a curved space-time with metric minimally coupled to the electromagnetic field
we obtain from $\cs$ the following equations
\begin{eqnarray}\label{EQPV}
F^{\mu \nu}_{\; ; \mu} + \frac {\sigma} {c} (A^{\mu} v^{\nu} -  
A^{\nu} v^{\mu})_{\; ; \mu}  =  J^{\nu}
\label{equ}
\end{eqnarray}
where the semicolon denotes the covariant derivative, 
$J^{\nu}= (-\sigma /c)v_{\mu}F^{\mu\nu}$ and
$v_{\mu} = R(t)\,\delta ^{0}_\mu $ the four-velocity of 
the \ae ther.

Here we notice that our new equations (\ref{equ}) differ from the ones
of our previous model \cite{Alex3} by the presence of a skew-symmetric 
term $(A^{\mu} v^{\nu} -  A^{\nu} v^{\mu})_{\; ; \mu}$ instead of the term 
$(1 / \lambda ^2){A}^{\nu}$ (that comes from the Proca term). In both cases,
the qualitative behaviour of the new term will be very similar to the one
obtained in the Proca model as we can see after comparing Table 3 
below and the graphics of our preceding paper.

For $\nu = 2$ we have 
\begin{eqnarray} \label{EQPf}
\ddot{f} + \left[4\;k_{c} -  
 \frac{\sigma}{c}\; \dot{R} \right]\,f  = 0
\end{eqnarray}
\noindent
and according to the geometry, we have to solve the equations below:
\begin{eqnarray} \label{EQG}
\ddot{f} - \alpha t \;\frac{\sigma}{c}\;f &= &0 \;\;\; \mbox{for $k_{c}=0$} \; ;\nonumber \\ 
\ddot{f} + (4 - \;\frac{\sigma}{c} \alpha \sin t) \;f &= &0  \;\;\; \mbox{for $k_{c}=+1$} \; ;\\
\ddot{f} - (4 + \;\frac{\sigma}{c} \alpha \sinh t) \;f &= &0  \;\;\; \mbox{for $k_{c}= -1$} \;.\nonumber
\end{eqnarray}
We assume an initial field of magnitude 
${\cal B}(t_{i}) \approx 10^{-21}$T. Since our model should also contain 
a weak initial electric field to be dissipated during the evolution of Universe, 
we shall assume the existence of an electric field of 
magnitude ${\cal E}(t_{i}) \approx 10^{-4}$V/m. These limits are fixed in order to
give, as a final result, a realistic value for the modulus of ${\cal B}(t_{f})$  
that agrees with the one established by the usual theory of the cosmic fields. 
These initial values we have assumed don't perturb the gravitational field, as it is evident
from simple calculations which show the energy-momentum tensor of the electromagnetic 
and gravitational field related by a factor above $10^{10}$.
These results then justify the minimal coupling of these fields as a suitable
framework to study the influence of the very strong geometry of the Universe 
on the very weak electromagnetic phenomenon.

Using the same method employed in \cite{Alex3}, we will integrate 
numerically equations (\ref{EQG}) from the initial cosmic conformal 
time $ t_{i} = 0.0890$ to the final time $t_{f} = 1.6100$. 
In standard cosmology this range corresponds respectively to the 
final stage of the matter-radiation coupling and our current epoch. 
We have obtained around $20,000$ points in the numerical integration 
of which, in Table 2, we displayed only a small ensemble of these points 
that can give us a qualitative view of the amplification phenomenon.
\begin{table}[h]
\caption{Some Numerical Data (F=Flat, E=Elliptic, H=Hyperbolic)}
\label{tabela3}
\begin{center}
\begin{tabular}{c||c|c|c|c|c|c} 
\hline
      t     & 0.0890 & 0.1000& 0.2352 & 0.5779 & 1.5793 & 1.6100
  \\  \hline    \hline  
(F)  $\;log|{\cal E}|$&-3.0004 &-3.1804&-4.7392&-6.2559 &-8.0335 &-8.0147           
  \\  \cline{2-7}   
\hspace{0.5cm} $\;log|c{\cal B}|$&-11.6993 &-4.7811&-5.2158&-6.2569 &-7.5110 &-7.5378           
 \\ \hline  \hline 
(E) $\;log|{\cal E}|$&-2.9997&-3.2035&-4.7034&-6.4784 &-7.8183 &-7.8333           
  \\  \cline{2-7}   
 \hspace{0.5cm}$\;log|c{\cal B}|$&-11.6988 &-4.8575&-5.2248&-6.3072 &-8.6010 &-8.7447           
 \\ \hline  \hline 
(H) $\;log|{\cal E}|$&-3.0009 &-3.2052&-4.6926&-6.2743&-8.1737 &-8.2008           
  \\  \cline{2-7}   
 \hspace{0.5cm}$\;log|c{\cal B}|$&-11.6999 &-4.8591&-5.2266&-6.2841&-7.6994 &-7.7205           
 \\ \hline  
\end{tabular}
\end{center}
\end{table}
\noindent
These results are very similar to the ones of our latter work \cite{Alex3}.
In Table 3 we compared the initial and the final value of the fields 
for each geometry. The data show the important amplification 
of the ${\cal B}$ field, which is very welcomed by astrophysicists,
and the overall reduction of the electric field. It should be noticed 
that these results are determined not only by the evolution of the 
function $f(t)$, (which constrains the field equations) but also by the 
direct contribution of the geometry as given by the scale factor, $R(t)$, that is 
present in both mathematical expressions for ${\cal E}$ and ${\cal B}$. 
It is the interchange between the gravitational 
and electromagnetic fields that imposes, as the Universe evolves, 
the decreasing of the electric field and the amplification of the ${\cal B}$ 
field.
\begin{table}[h]
\caption{The Reduction of ${\cal E}$ and the Amplification of ${\cal B}$}
\label{tabela2}
\begin{center}
\begin{tabular}{c c c} \hline
              Geometry  &  ${\cal E}$ & ${\cal B}$    
  \\ \hline 
   Flat ($k_{c}=0$)       & $\times 10^{-5}$  &   $\times 10^{+4}$           
  \\  
  Elliptic, ($k_{c}=+1$) &  $\times 10^{-5}$      &   $\times 10^{+3}$
 \\   
Hyperbolic, ($k_{c}=-1$) &  $\times 10^{-5}$  & $\times 10^{+4}$
 \\ \hline
\end{tabular}
\end{center}
\end{table}

{\raggedright
\section{Conclusion} }  \label{I}

Our model for the \AE ther of Dirac allow us to adapt our description
to any observer, be it inertial or not. In the case of an observer at rest relative
to the \ae ther, in a flat spacetime, we have obtained the same description 
of previous models \cite{Vigier, Kar} but with different material relations
involving the fields $\vec{D},\vec{H},\vec{E},\vec{B}$. This accounts for the
fact that although the \ae ther is a medium with conductivity, the presence of 
an electromagnetic field disturbs its isotropy. The equations 
we have obtained in this case had the same form as the Maxwell equations in a 
conductive medium provided we adopted a certain ``gauge'' choice for the vector 
$\vec{K}$ (\ref{K1},\ref{K3}). Here we notice the same role of the gauge condition as in the
original work of Dirac's NE \cite{Dirac2}, e.g. as a condition that determines a certain physics.
In the case of a curved background, an observer with $v_{\alpha} = R(t) \delta^{0}_{\alpha}$ 
will see the same phenomenon of amplification of the magnetic induction and reduction
of the electric intensity that had already been observed in \cite{Alex3} in the 
context of a Proca electromagnetic field in a Dirac \ae ther. 

The material relations between the polarization tensor and 
the gauge potential given in (\ref{mat3},\ref{mat6}) shows that 
the tensor $H^{\mu\nu}$ is related to the electromagnetic 
properties of the medium (encoded into the tensor $S^{\mu}_{\a\b}$) 
and the metric tensor. This raises an interesting picture of a mutual 
effect between the electromagnetic field and the geometry of the spacetime. 
In fact, if the GA describes how the expansion of the universe influences 
the electromagnetic fields as a kind of geometric background effect
that should be added to well-established non-geometric effects, there is
also the possibility for the electromagnetic field to influence the 
expansion of the universe \cite{Tsagas}. 

The geometric relations between the electromagnetic field 
and the metric tensor involves the scale factor $R(t)$. 
The amplification of the field ${\cal B}$ in this case is determined
by the minimal coupling we adopted. In other coulplings between 
the electromagnetic and the gravitational fields the interchange 
could be more rapid and/or more intense, as we can see in \cite{Turner}. 
Our results confirm once more the strict relations between the 
electromagnetic and gravitational phenomena. Furthermore, the model we elaborated 
for the Dirac's \AE ther in a curved space-time sugests it can also be 
applied to the study of the electromagnetism in the large scale structure of the Universe. 



\begin{thebibliography}{99}
\bibitem{Dirac2} P. A. M. Dirac, {\em Proc. Roy. Soc.\ A \/} 
{\bf 209} (1951) 291-296;

\bibitem{Dirac} P. A. M. Dirac, {\em Nature} {\bf 168} (1951) 906-907;

\bibitem{Vigier} J. -P. Vigier, {\em IEEE Trans. Plasma Sci. \/} {\bf 18 } (1990) 64;
 
\bibitem{Kar} G. Kar, M. Sinha e S. Roy, 
{\em Int. J. Theor. Phys. \ \/} {\bf 32} (1993) 593-607;

\bibitem{Alex3} A. L Oliveira,  {\em Mod. Phys. Lett. \ A \/} 
{\bf 16} (2001) 541-555;

\bibitem{Sudarshan} K. P. Sinha, C. Sivaram and 
E. C. G. Sudarshan,  {\em Found. Phys. \  \/} {\bf 6} (1976) 65-70;

\bibitem{Lehnert} B. Lehnert,  {\em Found. Phys. Lett \  \/} {\bf 15} (2002) 95;

\bibitem{Schwinger} J. Schwinger, 'A Report on Quantum Electrodynamics`, 
in "The Physicist's Conception of Nature", Ed. J. Mehra, Reidel, Dordrecht, 1973;


\bibitem{volkov} A. M. Volkov, V. A. Kiselev, {\em Sov. Phys. Jetp}  {\bf 30} (1970) 733; 

\bibitem{plebanski} J. Plebanski, {\em Phys. Rev.} {\bf 118} (1960) 1396;

\bibitem{landau} L. D. Landau and E. M. Lifshitz , "The Th\'eorie du Champ",  \'Editions MIR, Moscou, 1966;

\bibitem{Rubinstein} D. Grasso and H. R. Rubinstein, 
{\em Phys. Rep.} {\bf 348} (2001) 163-266; Magnetic Fields in the Universe; 

\bibitem{Parker} E. N. Parker , "Cosmical Magnetic Fields", 
Clarendon Press, Oxford, 1979;

\bibitem{Tsagas} D. R. Matravers and C. G. Tsagas, 
{\em Phys. Rev. \ D \/} {\bf 62} (2000) 103519; C. G. Tsagas, 
{\em Phys. Rev. Lett. \  \/} {\bf 86} (2001) 5421-5424;

\bibitem{Turner} M. S. Turner e L. M. Widrow, {\em Phys. Rev.\ D \/} 
{\bf 37} (1988) 2743-2754; M. Hindmarsh e A. Everett, {\em Phys. Rev.\ D \/} 
{\bf 58} (1998) 103505; M. Christensson e M. Hindmarsh, 
{\em Phys. Rev.\ D \/} {\bf 60} (1999) 063001;


\end{thebibliography}
\end{document}